\documentclass[prd,twocolumn,showpacs,floatfix,superscriptaddress]{revtex4}

 \pdfoutput=1

\usepackage[utf8]{inputenc}
\usepackage{graphicx}
\usepackage{epsfig}
\usepackage{bm}
\usepackage{amssymb}
\usepackage{float}
\usepackage{amsmath}
\usepackage{dcolumn}
\usepackage{cancel}
\usepackage{graphicx}
\usepackage{epsfig}
\usepackage{bm}
\usepackage[greek,english]{babel} 
\usepackage{alphabeta}
\usepackage{amssymb}
\usepackage{float}
\usepackage{amsmath}
\usepackage{dcolumn}
\usepackage{cancel}
\usepackage[colorlinks]{hyperref}
\usepackage[usenames,dvipsnames]{color}
\hypersetup{
     breaklinks=true,
    pdfstartview={FitH},    
    colorlinks=true,       
    linkcolor=blue,          
    citecolor=red,        
    filecolor=magenta,      
    urlcolor=blue,           
    anchorcolor=green,      
    linktocpage=true
}

\newcommand{\bea}{\begin{eqnarray}}
 
\newcommand{\eea}{\end{eqnarray}}
\newcommand{\bma}{\begin{pmatrix}}
\newcommand{\ema}{\end{pmatrix}}
\newcommand{\be}{\begin{equation}}
\newcommand{\ee}{\end{equation}}
\newcommand{\beno}{\begin{equation*}}
\newcommand{\eeno}{\end{equation*}}

\renewcommand{\k}{\vec{k}}

\newcommand{\nn}{\nonumber}

\def\doi{http://doi.org}
\def\arxiv{http://arxiv.org/abs}

\def\a{\alpha}
\def\b{\beta}

\def\d{\delta}

\def\la{\lambda}

\def\k{\kappa}
\def\m{\mu}
\def\n{\nu}

\def\f{\phi}

\def\z{\zeta}

\begin{document}

\title{Alleviating $H_0$ Tension with New Gravitational Scalar Tensor  
Theories}

\author{Shreya Banerjee}
\email{shreya.banerjee@fau.de}
\affiliation{\mbox{Institute for Quantum Gravity, FAU Erlangen-Nuremberg, Staudtstr. 7, 91058 Erlangen, Germany}}

\author{Maria Petronikolou}
\email{petronikoloumaria@mail.ntua.gr}
\affiliation{ \mbox{National Observatory of Athens, Lofos Nymfon, 11852 Athens, Greece} }

\author{Emmanuel N. Saridakis}
\email{msaridak@noa.gr}
\affiliation{ \mbox{National Observatory of Athens, Lofos Nymfon, 11852 Athens, Greece} }
\affiliation{CAS Key Laboratory for Researches in Galaxies and Cosmology, 
Department of Astronomy, \\
University of Science and Technology of China, Hefei, 
Anhui 230026, P.R. China}
\affiliation{\mbox{Departamento de Matem\'{a}ticas, Universidad Cat\'{o}lica del 
Norte, 
Avda.
Angamos 0610, Casilla 1280 Antofagasta, Chile}}

\begin{abstract} 
We investigate the cosmological  applications of new gravitational scalar-tensor 
theories and we analyze them in the light of $H_0$ tension. In these theories 
the Lagrangian contains the Ricci scalar and its first and second derivatives 
in a specific combination that makes them free of ghosts, thus corresponding 
to healthy bi-scalar extensions of general relativity.  
We examine two specific models,  and for particular choices of the model 
parameters  we find that the effect of the additional terms  is negligible at 
high redshifts, obtaining a coincidence with   $\Lambda$CDM cosmology, however 
as time passes the deviation increases and thus at low redshifts the Hubble 
parameter acquires increased values ($H_0\approx 74 km/s/Mpc$) in a controlled 
way. The mechanism behind 
this behavior is the fact that the effective dark-energy equation-of-state 
parameter exhibits phantom behavior,  which implies 
faster expansion, which is one of the {\bf{sufficient}} conditions that are 
capable of alleviating the $H_0$ tension.
 Lastly,   we 
confront the   models with   Cosmic Chronometer (CC) data showing full 
agreement within 1$\sigma$ confidence level.
\end{abstract}


\maketitle

\section{Introduction}

Although the concordance  $\Lambda$CDM  paradigm is very successful 
in describing early- and late-time cosmological evolution at both background 
and perturbation levels, nevertheless the last years there have appeared some 
potential tensions with specific datasets, such as the  $H_0$ and $\sigma_8$ 
ones. In particular,    the estimation for the present Hubble parameter $H_0$ 
according to the Planck collaboration  and assuming  $\Lambda$CDM  scenario 
is $H_0 = (67.27\pm 0.60)$ km/s/Mpc 
\cite{Aghanim:2018eyx}, which is 
 in tension at about $4.4\sigma$ with the direct 
measurement of the  $2019$ SH0ES collaboration (R19), namely 
$H_0 = (74.03\pm1.42)$ km/s/Mpc, obtained using long-period Cepheids 
\cite{Riess:2019cxk}. On the other hand, the $\sigma_8$ 
tension arises from the fact that the   parameter  that 
quantifies the  matter clustering within spheres of radius $8 h^{-1} 
\text{Mpc}$, is found to be different from   Cosmic Microwave 
Background (CMB) estimation 
\cite{Aghanim:2018eyx} and from SDSS/BOSS measurement
\cite{Zarrouk:2018vwy, Alam:2016hwk, Ata:2017dya}.
These tensions, and especially the $H_0$ one,  progressively seem not to be 
related to  unknown systematics, opening the road to many modifications of the 
standard lore \cite{DiValentino:2020zio,DiValentino:2020vvd} (for a review 
see \cite{Abdalla:2022yfr}).

One may follow two main ways to alleviate the $H_0$ tension. The first is to 
modify  
the  universe content and/or particle interactions while 
keeping general relativity as the underlying gravitational theory     
\cite{DiValentino:2021izs, DiValentino:2015ola,  
Bernal:2016gxb,Kumar:2016zpg, 
DiValentino:2017iww,  DiValentino:2017oaw,  Binder:2017lkj, 
DiValentino:2017zyq,Yang:2018euj, 
DEramo:2018vss,Poulin:2018cxd,Yang:2018qmz,Yang:2018prh,Pan:2019gop, 
Pan:2019jqh, 
Shafieloo:2016bpk,
 Pandey:2019plg,Adhikari:2019fvb, 
Benisty:2019pxb,
 Perez:2020cwa,Pan:2020bur, 
Benevento:2020fev,Banerjee:2020xcn,Elizalde:2020mfs,Alvarez:2020xmk,
DeFelice:2020cpt, Haridasu:2020pms,
Seto:2021xua,Bernal:2021yli, 
Alestas:2021xes,Elizalde:2021kmo,Krishnan:2021dyb,Theodoropoulos:2021hkk}.
The second way is to construct gravitational modifications,
which applied to cosmological framework would lead to altered expansion rate
  \cite{Hu:2015rva,Khosravi:2017hfi,Belgacem:2017cqo,El-Zant:2018bsc,
Basilakos:2018arq,Adil:2021zxp,Nunes:2018xbm,Lin:2018nxe, 
Escamilla-Rivera:2019ulu,DiValentino:2019jae,Vagnozzi:2019ezj, 
Anagnostopoulos:2019miu, Cai:2019bdh,Braglia:2020auw,
Anagnostopoulos:2020lec,Capozziello:2020nyq,Saridakis:2019qwt, 
DAgostino:2020dhv,Abadi:2020hbr,Barker:2020gcp, Wang:2020zfv,Ballardini:2020iws,
LinaresCedeno:2020uxx,daSilva:2020bdc,Odintsov:2020qzd,Nojiri:2022ski,Nojiri:2019fft}. We mention here that 
modified gravity has additional advantages too, such as the improvement of the 
renormalizability behavior of general relativity as well as the description of 
inflationary and/or dark-energy phases, and thus it might be more preferable. 
Finally, there is another way to alleviate $H_0$ tension, in the framework 
of the running vacuum models \cite{Sola:2017znb}, based  
on quantum field
theory in curved spacetime 
\cite{SolaPeracaula:2021gxi,Moreno-Pulido:2022phq,Moreno-Pulido:2020anb}, 
without the need to acquire phantom behavior
(for a review of both the theoretical and phenomenological situation  see 
\cite{SolaPeracaula:2022hpd} and references therein).

In the present work we are interested in alleviating the $H_0$ tension in the 
framework of new gravitational scalar-tensor theories
\cite{Naruko:2015zze, Saridakis:2016ahq,Saridakis:2018fth}. In such 
constructions one  uses  Lagrangians with the Ricci scalar as well as its 
first and second derivatives, nevertheless in combinations that result to 
ghost-free theories. These theories are found to have $2+2$ propagating degrees 
of freedom, and thus  falling outside Horndeski/Galileon 
\cite{Horndeski:1974wa,Nicolis:2008in,Deffayet:2011gz} and beyond-Horndeski 
theories \cite{Gleyzes:2014dya}. However, although they are bi-scalar 
extensions of general relativity, they were named ``new gravitational 
scalar-tensor theories'' since
they can still be expressed in pure 
geometrical terms \cite{Naruko:2015zze}.

 The plan of the work is the following: In Section \ref{Overview} we briefly 
review the  new gravitational 
scalar-tensor theories, and in Section \ref{CosmBehaviour} we apply them to a 
cosmological framework, extracting the modified Friedmann equations. Then, in 
Section \ref{HubbleTension}  we construct specific models
 that can alleviate the  $H_0$ tension, and we compare the induced behavior  to 
that of $\Lambda$CDM scenario as well as to Cosmic Chromometers (CC) data. 
Finally, in Section 
\ref{Conclusions} we provide the conclusions.

\section{Overview}
\label{Overview}

In this section we give a brief overview of the gravitational scalar-tensor 
theories.  
The action of such constructions is   given as \cite{Naruko:2015zze, 
Saridakis:2016ahq}
\begin{equation}
\label{bfR}
S=\int d^{4}\sqrt{-g}\, f\left(R,(\nabla R)^{2},\square R \right),
\end{equation}
with $(\nabla R)^{2}=g^{\m\n}\nabla_{\m}R\nabla_{\n}R$. In the following 
we 
set the Planck mass $M_{P}=1/\kappa=1$, where $\kappa$ is the gravitational 
constant, for simplicity. 
One can   rewrite the above action  by converting   the Lagrangian  using 
double Lagrange multipliers, resulting to actions of multi-scalar fields 
coupled minimally to gravity. In order to achieve it, one fixes the 
dependence of $f$ on $\square R=\beta$.

In the present work, we consider theories with the following    $f$ form:
\begin{equation}
f(R,(\nabla R)^{2},\square R)=\mathcal{K}((R,(\nabla R)^{2})+\mathcal{G}(R,(\nabla 
R)^{2})\square R,
\end{equation}
thus maintaining a linear form in    $\square R=\beta$. Generalizations to 
non-linear forms are straightforward, although more complicated.
In this case,    (\ref{bfR}) transforms to   
\begin{eqnarray}
\label{action}
&&\!\!\!\!\!\!\!\!\!\!\!\!\!\!\!
S=\int d^{4}x 
\sqrt{-\hat{g}}\left[\frac{1}{2}\hat{R}-\frac{1}{2}\hat{g}^{\m\n}\nabla_{\m}\chi\nabla_{\n
}\chi\right.\nonumber\\
&& \ \ \ \ \ \ \ \ \ \ \ \
- 
\frac{1}{\sqrt{6}}e^{-\sqrt{\frac{2}{3}}\chi}\hat{g}^{\m\n}\mathcal{G}\nabla_{\m}
\chi\nabla_{\n}\phi+\frac{1}{4}e^{
-2\sqrt{
\frac{2}{3}}
\chi}\mathcal{K}
\nonumber\\
&& \ \ \ \ \ \ \ \ \ \ \ \
\left.
+\frac{1}{2}e^{-\sqrt{\frac{2}{3}}\chi}\mathcal{G}\hat{\square}\phi-\frac{1}{4}e^{-\sqrt{
\frac{2}{3}}\chi}\phi\right],
\end{eqnarray}
where  
$
\mathcal{K}=\mathcal{K}(\phi,B)$ and $\mathcal{G}=\mathcal{G}(\phi,B),
$
 with
$
B=2e^{\sqrt{
\frac{2}{3}}\chi}g^{\m\n}\nabla_{\m}\phi\nabla_{\n}\phi.
$
 The  $\chi$ and $\phi$ fields  are introduced through the   conformal 
transformations  
$g_{\mu\nu}=\frac{1}{2}e^{-\sqrt{\frac{2}{3}}\chi}\hat{g}_{\mu\nu}$,
$\varphi\equiv f_{\b}$, and they enter  in a specific combination in a way that 
the final form of the action is equivalent to the original higher-derivative 
gravitational action.

Varying the action (\ref{action}) with respect to the metric leads to  the 
following field 
equations in Einstein frame \cite{Naruko:2015zze, 
Saridakis:2016ahq}:
\begin{eqnarray}
\label{metricfieldeq}
\mathcal{E}_{\m\n}&\!\!=\!\!&\frac{1}{2}G_{\m\n}+\frac{1}{4}g_{\m\n}g^{\a\b}\nabla_{\a}
\chi\nabla_
{\b}\chi-\frac{1}{2}\nabla_{\m}\chi\nabla_{\n}\chi\nonumber\\
&&
+\frac{1}{4}g_{\m\n}\sqrt{\frac{2}{3}}e^
{-\sqrt{ \frac{2}{3}}
\chi}g^{\a\b}\mathcal{G}\nabla_{\a}\chi\nabla_{\b}\phi\nonumber\\
&&
-\frac{1}{2}\sqrt{\frac{2}{3}}e^{
-\sqrt{\frac{
2}{3}}\chi}\mathcal{G}\nabla_{(\m}\chi\nabla_{\n)}\phi \nn\\
&&-\sqrt{\frac{2}{3}}g^{\a\b}\nabla_{\a}\chi\nabla_{\b}\phi\,\mathcal{G}_{B}\nabla_{\m}
\phi\nabla_{\n}\phi
\nonumber\\
&&
-\frac{1}{4}g_{\m\n}e^{-\sqrt{\frac{2}{3}}\chi}\mathcal{G}
\square\phi+\mathcal{G}_{B 
}(\square\phi)\nabla_{\m}\phi\nabla_{\n}\phi
\nonumber\\
&&
+\frac{1}{2}e^{-\sqrt{\frac{2}{3}}\chi}
\mathcal{G} \nabla_{\m}\nabla_{\n}\phi 
-\frac{1}{2}\nabla_{\k}\left(e^{-\sqrt{\frac{2}{3}}\chi}\mathcal{G}\d^{\la}_{(\m}\d^{\k}
_{\n)} 
\nabla_{\la}\phi\right)
\nonumber\\
&&
+\frac{1}{4}\nabla_{\k}\left(e^{-\sqrt{\frac{2}{3}}\chi}\mathcal{G}
g_{\m\n}\nabla^{\k}\phi\right)-\frac{1}{8}g_{\m\n}e^{-2\sqrt{\frac{2}{3}}\chi}\mathcal{K}
\nonumber\\
&&
+\frac{1}{ 2}e^{-\sqrt{\frac{2}{3}}\chi}\mathcal{K}_{B}\nabla_{\m}\phi\nabla_{\n}\phi 
+\frac{1}{8}g_{\m\n}e^{-\sqrt{\frac{2}{3}}\chi}\phi=0.
\end{eqnarray}
Additionally, varying (\ref{action}) with respect to  $\chi$ 
and $\phi$ gives rise to field equations as 
\begin{eqnarray}
\mathcal{E}_{\chi}&\!\!=\!\!&\square\chi+\frac{1}{3}e^{-\sqrt{\frac{2}{3}}\chi}g^{\m\n}
\mathcal{ G}
\nabla_{\m}\chi\nabla_{\n}\phi\nonumber\\
&&
-\frac{2}{3}g^{\m\n}\nabla_{\m}\chi\nabla_{\n}\phi\,\mathcal
{G}_{B}g^ 
{\a\b}\nabla_{\a}\phi\nabla_{\b}\phi\nonumber\\
&&
+\frac{1}{2}\sqrt{\frac{2}{3}}\nabla_{\m}\left( 
e^{-\sqrt{\frac{2}{3}}\chi}g^{\m\n}\mathcal{G}\nabla_{\n}\phi\right)\nn\\
&&-\frac{1}{2}\sqrt{\frac{2}{3}}e^{-\sqrt{\frac{2}{3}}\chi}\mathcal{G}\square\phi+\sqrt{
\frac{2}{3}}
\mathcal{G}_{B}\nabla_{\m}\phi\nabla_{\n}\phi\,g^{\m\n}\square\phi
\nonumber\\
&&
-\frac{1}{2}\sqrt{\frac{
2}{3}}e^{-
2\sqrt{\frac{2}{3}}\chi}\mathcal{K}+\frac{1}{2}e^{-\sqrt{\frac{2}{3}}\chi}\mathcal{K}_{B}
\sqrt{\frac{2}{3}}g^{\m\n}\nabla_{\m}\phi\nabla_{\n}\phi
\nn\\
&&+\frac{1}{4}\sqrt{\frac{2}{3}}e^{-\sqrt{\frac{2}{3}}\chi}\phi=0,
\label{chifieldeq}
\end{eqnarray}
and 
\begin{eqnarray}
\mathcal{E}_{\phi}&\!\!=\!\!&-\frac{1}{2}\sqrt{\frac{2}{3}}e^{-\sqrt{\frac{2}{3}}\chi}g^{
\m\n }
\mathcal{G}_{\phi}\nabla_{\m}\chi\nabla_{\n}\phi
\nonumber\\
&&
+2\sqrt{\frac{2}{3}}\nabla_{\b}\left(g^{
\m\n}\mathcal{G} 
_{B}g^{\a\b}\nabla_{\a}\phi\nabla_{\m}\chi\nabla_{\n}\phi\right)
\nonumber\\
&&
+\frac{1}{2}\sqrt{\frac{2}
{3}} 
\nabla_{\n}\left(e^{-\sqrt{\frac{2}{3}}\chi}g^{\m\n}\mathcal{G}\nabla_{\m}\chi\right)\nn\\
&&+\frac{1}{2}e^{-\sqrt{\frac{2}{3}}\chi}\mathcal{G}_{\phi}\square\phi-2\mathcal{G}_{B}
(\square\phi)
^{2}-2\nabla_{\n}\mathcal{G}_{B}\square\phi\nabla^{\n}\phi
\nonumber\\
&&
-\frac{1}{2}\sqrt{\frac{2}{3}}
\nabla^{\m}\left(e^{-\sqrt{\frac{2}{3}}\chi}\nabla_{\m}\chi\,\mathcal{G} 
\right)+\frac{1}{2}\nabla^{\m}\left(e^{
-\sqrt{\frac{2}{3}}\chi}\,\mathcal{G}_{\phi}\nabla_{\m}\phi\right)\nn\\
&&-\frac{1}{2}\sqrt{\frac{2}{3}}e^{-\sqrt{\frac{2}{3}}\chi}\nabla^{\m}\chi\mathcal{G}_{B}
\nabla_{\m}
B+\frac{1}{2}e^{-\sqrt{\frac{2}{3}}\chi}\nabla^{\m}\mathcal{G}_{B}\nabla_{\m}B
\nonumber\\
&&
+\sqrt{\frac{2}{3}}
e^{-\sqrt{
\frac{2}{3}}\chi}\mathcal{G}_{B}\nabla^{\m}\left(e^{\sqrt{\frac{2}{3}}
\chi}\nabla_{\m}\chi\nabla^{\n}\phi\nabla_{\n}\phi\right)\nn\\
&&
+2 e^{-\sqrt{\frac{2}{3}}\chi}\mathcal{G}_{B}  
\nabla^{\m}\left(e^{\sqrt{\frac{2}{3}}\chi}\nabla^{
\n}\phi\right)\nabla_{\m}\nabla_{\n}\phi
\nonumber\\
&&
+2\mathcal{G}_{B}R^{\m\n}\nabla_{\m}\phi\nabla_{\n
}\phi+\frac{1}{4}e^{-2\sqrt{\frac{2}{3}}\chi}\mathcal{K}_{\phi}\nn\\
&&-\nabla_{\n}\left(e^{-\sqrt{\frac{2}{3}}\chi}\mathcal{K}_{B}g^{\m\n}\nabla_{\m}
\phi\right)-\frac{
1}{4}e^{-\sqrt{\frac{2}{3}}\chi}=0,
\label{phifieldeq}
\end{eqnarray}
where for simplicity we have neglected the hats.
Here, the subscripts in $\mathcal{G}$ and $\mathcal{K}$ denote the partial derivatives and the symmetrization is indicated by the parentheses in spacetime indices.
The above equations reduce to GR for $\mathcal{K}=\phi/2$ and $\mathcal{G}=0$, 
with the conformal transformation in this case  being
$\chi=-\sqrt{\frac{3}{2}}\ln2$. As we can see, 
 the above equations do not contain any higher derivative terms, and  therefore 
the present theory is well-behaved.   Lastly, note that since we have set 
the Planck mass to one, the field $\chi$ is dimensionless while 
$\phi$ has   dimensions of $[M]^2$.

 \section{Cosmological Behaviour}
 \label{CosmBehaviour}

 We can now proceed to the  study of the cosmological behaviour of the present 
model. For this we consider a flat Friedmann-Robertson-Walker (FRW) metric
 \be
ds^{2}=-dt^{2}+a(t)^{2} \delta_{ij}dx^{i}dx^{j},
\ee
with $a(t)$   the scale factor. We further assume that the two scalars are 
time-dependent only.

Including the matter sector, considered to correspond to a perfect fluid, the 
metric  
field equations (\ref{metricfieldeq})   become
\be
\mathcal{E}_{\m\n}=\frac{1}{2}T_{\m\n},
\ee
with $T_{\m\n}=\frac{-2}{\sqrt{-g}}\frac{\d S_{m}}{\d g^{\m\n}}$ representing 
the matter energy-momentum 
tensor. 

With the above substitutions into equations (\ref{metricfieldeq}), we obtain 
the following  
Friedmann equations:
\begin{eqnarray}
&&\!\!\!\!\!\!\!\!\!\!\!\!\!\!\!\!\!
3H^{2}-\rho_m-\frac{1}{2}\dot{\chi}^{2}+\frac{1}{4}e^{-2\sqrt{\frac{2}{
3}}\chi}\mathcal{K}
\nonumber\\
&& \!\!\!\!
+\frac{2}{3}\dot{\phi}^{2}\left[\dot{\phi}\left(\sqrt{6}\dot{\chi}
-9H\right)-3\ddot{\phi} 
\right]\mathcal{G}_{B}
\nn\\
&&  \!\!\!\!
-\frac{1}{2}e^{-\sqrt{\frac{2}{3}}\chi}\left[\dot{B}\dot{\f}\mathcal{
G}_{B}+ \frac{\f}{2} +\dot{\f}^{2}\left(\mathcal{G}_{\f}-2\mathcal{K}_{B} \right) 
 \right]\! =0,  
\label{FR1}
\end{eqnarray}
\begin{eqnarray}
&&\!\!\!\!\!\!\!\!\!\!\!\!\!\!\!\!\!\!\!\!\!\!
3H^{2}+2\dot{H}+p_m+\frac{1}{2}\dot{\chi}^{2}+\frac{1}{4}e^{-2\sqrt{
\frac{2}{3}}
\chi}\mathcal{K}
\nonumber\\
&&\!\!\!\!\!\!\!\!\!
+\frac{1}{2} e^{-\sqrt{\frac{2}{3}}\chi}\left(-\frac{\f}{2}+\dot{B}\dot{\f
}\mathcal{
G}_{B}+\dot{\f}^{2}\mathcal{G}_{\f} \right)=0,
  \label{FR2}
\end{eqnarray}
with $B(t)=2 e^{\sqrt{\frac{2}{3}}\chi} g^{\m\n}\nabla_{\m}\f\nabla_{\n}\f=-2 
e^{\sqrt{\frac{2}{3}}\chi} \dot{\f}^{2}$, and  $H=\dot{a}/a$   the Hubble 
parameter, where dots denoting differentiation with respect to $t$.
Similarly, the two scalar field equations (\ref{chifieldeq}) and 
(\ref{phifieldeq}) 
lead to:
\begin{eqnarray}
&&\!\!\!\!\!\!\!\!\!\!\!\!\!\!
\mathcal{E}_{\chi}=\ddot{\chi}+3H\dot{\chi}-\frac{1}{3}\dot{\f}^{2}\left[\dot{\f}
\left(3\sqrt{6}H-
2\dot{\chi} \right)+\sqrt{6}\ddot{\f} \right]\mathcal{G}_{B}\nn\\
&&+
\frac{1}{
2\sqrt{6}}
e^{-\sqrt{\frac{2}{3}}\chi}\left[2\dot{B}\dot{\f}\mathcal{G}_{B}-\f+2\dot{\f}^{2}
\left(\mathcal{K}_{
B}+\mathcal{G}_{\f} \right) \right]\nonumber\\
&&+\frac{1}{\sqrt{6}}e^{-2\sqrt{\frac{2}{3}}\chi}\mathcal{K}
=0,
\label{chiequation}
\end{eqnarray}
and
\begin{eqnarray}
&&\!\!\!\!\!\!\!\!\! \!
\mathcal{E}_{\phi}=
\frac{1}{3}e^{-\sqrt{\frac{2}{3}}\chi}
\left[\dot{\f}\left(-9H+\sqrt{6}\dot{\chi}
\right)-3\ddot{\f}\right]\mathcal{K}_{B}
\nn
\\
&&
+\frac{1}{6}\dot{B}\left\{3e^{-\sqrt{\frac{2}{3}}
\chi}\dot{B} +4\dot{\f}
\left[\dot{\f}\left(9H-\sqrt{6}\dot{\chi}\right)+3\ddot{\f}\right]\right\}\mathcal{G}_{BB}
\nn\\
&&
+\frac{1}{3}e^{-\sqrt{
\frac{2}{3}
}\chi}\left[\dot{\f}\left(9H-\sqrt{6}\dot{\chi}\right)+3\ddot{\f}\right]\mathcal{G}_{\f}
\nn
\\
&&
+\left\{e^{-\sqrt{\frac{2}{3}}\chi}\dot{B}\dot{\f}+\frac{2}{3}\dot{\f}^{2}
\left[\dot{\f}\left(9H-\sqrt{6}\dot{\chi}\right)+3\ddot{\f}\right]\right\}\mathcal{G}_{B 
\f}
\nn
\\
&&
-e^{-\sqrt{\frac{2}{3}}\chi}\dot{\f}^{2}\mathcal{K}_{B 
\f}+\frac{1}{2}e^{-\sqrt{\frac{2}{3}}\chi}\dot{\f}^{2}\mathcal{G}_{\f\f}
 -e^{-\sqrt{\frac{2}{3}}\chi}\dot{B}\dot{\f}\mathcal{K}_{BB}
 \nn
\\
&&
+\left[
 \frac{4}{3}\dot{\f}\left(9H-2\sqrt{6}\dot
{\chi} 
\right)\ddot{\f}
-\frac{1}{\sqrt{6}}e^{
-\sqrt{\frac{2}{3}}\chi}\dot{B}\dot{\chi}
\right.\nn\\
&&\left. \ \ \ \,
+\dot{\f}^2\left(18H^{2}+6\dot{H}-3\sqrt{6}H\dot{\chi}-\frac{2}{3}\dot{\chi}^{2}
-\sqrt{6} \ddot{\chi}\right)\right]
\mathcal{G}_{B}
\nn
\\
&&
-\frac{1}{4}
e^{-2\sqrt{\frac{2}{3}}\chi}\mathcal{K}_{\f}+ \frac{1}{4}e^{-\sqrt{\frac{2}{3}}\chi}
=0,
\label{phiequation}
\end{eqnarray}
with   $\mathcal{G}_{B \f}=\mathcal{G}_{\f 
B}\equiv\frac{\partial^{2}\mathcal{G}}{\partial B \partial \f}$, etc.

The above
Friedmann equations 
(\ref{FR1}),(\ref{FR2}) can be rewritten as
\begin{eqnarray}
&&H^2=\frac{1}{3}(\rho_{DE}+\rho_m)
 \label{FR1b}
 \\
&&2\dot{H}+3H^2=-(p_{DE}+p_m),
 \label{FR2b}
\end{eqnarray}
 with the effective dark energy and pressure defined as
\begin{eqnarray}
  \label{rhoDE}
 &&\!\!\!\!\!\!\!\!\!\!\!\!\!\!\!\!\!\!\!\!
 \rho_{DE}\equiv 
 \frac{1}{2}\dot{\chi}^{2}-\frac{1}{4}e^{-2\sqrt{\frac{2}{
3}}\chi}\mathcal{K}
\nn\\
&&\!\!
-\frac{2}{3}\dot{\phi}^{2}\left[\dot{\phi}\left(\sqrt{6}\dot{\chi}
-9H\right)-3\ddot{\phi} 
\right]\mathcal{G}_{B}
\nn\\
&&\!\!
+\frac{1}{2}e^{-\sqrt{\frac{2}{3}}\chi}\left[\dot{B}\dot{\f}\mathcal{
G}_{B}+ \frac{\f}{2} +\dot{\f}^{2}\left(\mathcal{G}_{\f}-2\mathcal{K}_{B} \right)
 \right],
 \end{eqnarray}
\begin{eqnarray}
\label{pDE}
&&\!\!\!\!\!\!\!\!\!\!\!\!\!\!\!\!\!\!\!\!\!\!\!\!\!\!\!\!\!\!\!\!\!\!\!\!\!\!\!\!\!
p_{DE}\equiv  
 \frac{1}{2}\dot{\chi}^{2}+\frac{1}{4}e^{-2\sqrt{
\frac{2}{3}}
\chi}\mathcal{K}
\nn\\
&&\!\!\!\!\!\!\!\!\!\!\!\!\!\!\!\!\!\!\!\!\!\!\!
+
\frac{1}{2} e^{-\sqrt{\frac{2}{3}}\chi}\left(\dot{B}\dot{\f
}\mathcal{
G}_{B}+\dot{\f}^{2}\mathcal{G}_{\f}-\frac{\f}{2} \right).
\end{eqnarray}
Hence, one can show that in the new gravitational scalar-tensor theories  the 
effective dark-energy  density satisfies
\begin{equation}
\dot{\rho}_{DE}+3H(\rho_{DE}+p_{DE})=0,
\end{equation}
while one can define the corresponding dark-energy equation-of-state parameter 
as
\begin{equation}
\label{wDE}
w_{DE}\equiv \frac{p_{DE}}{\rho_{DE}}.
\end{equation}

\section{Hubble Tension}
\label{HubbleTension}

In this section we construct specific models of the theory in order to be able 
to alleviate the $H_0$ tension. 
We mention here that   in modified gravity theories one  
typically has arbitrary functions, and thus she has a huge freedom to determine 
both their forms as well as their parameters. This freedom is similar to the 
freedom of choosing the arbitrary potentials in scalar-field cosmology. Hence, 
in the end of the day the obtained models are   phenomenological, aiming 
to be in agreement with observations. In the theories examined in the present 
manuscript, we 
consider specific ansatzes for the functions $\mathcal{K}(\phi,B)$ and 
$\mathcal{G}(\phi,B)$ and we select models that lead to higher Hubble function 
at low redshifts,  while introducing negligible deviations in the Hubble 
parameter at high redshifts as compared to $\Lambda$CDM. The two 
phenomenological models with the best behavior related to the $H_0$ tension are 
presented in the following.

\subsection{Model I}

As a first example we consider the following forms for $\mathcal{K}(\phi,B)$ and 
$\mathcal{G}(\phi,B)$:
\be
\mathcal{K}(\f,B)=\frac{1}{2}\f-\frac{\z}{2}B \quad \text{and} \quad 
\mathcal{G}(\f,B)=0,
\ee
with  $\zeta$   a coupling constant  with dimensions $[M]^{-4}$.
The corresponding Friedmann equations  (\ref{FR1}),(\ref{FR2}) read as
\begin{equation}
\label{Fr1rv}
3 
H^{2}-\rho_m-\frac{1}{2}\dot{\chi}^{2}+\frac{1}{8}e^{-2\sqrt{\frac{2}{3}}\chi}\f-\frac{1}{
4 }
e^{-\sqrt{\frac{2}{3}}\chi}\left(\f+\z\dot{\f}^{2} \right)=0,
\end{equation}
\bea
&&\!\!\!\!\!\!\!\!\!\!\!\!\!\!\!\!\!\!\!\!\!\!\!\!\!\!\!\!\!\!\!\!\!\!\!\!\!\!\!\!\!\!\!\!
\!
3 H^{2}+2 
\dot{H}+p_m+\frac{1}{2}\dot{\chi}^{2}+\frac{1}{8}e^{-2\sqrt{\frac{2}{3}}\chi}\f
\nonumber\\
&&\!\!
-\frac{1}{4}
e^{-\sqrt{\frac{2}{3}}\chi}\left(\f-\z\dot{\f}^{2} \right)=0,
\eea
while the two scalar field equations (\ref{chiequation}) and (\ref{phiequation}) become
\begin{equation}
\label{scalar1}
\ddot{\chi}+3 H 
\dot{\chi}+\frac{1}{2\sqrt{6}}e^{-2\sqrt{\frac{2}{3}}\chi}\f-\frac{1}{2\sqrt{6}}e^{-
\sqrt{\frac{2}{3}}\chi}\left(\f-\z\dot{\f}^{2} \right)=0,
\end{equation}
\begin{equation}
\label{scalar2}
\z \ddot{\phi}+\frac{1}{3}\z\dot{\f}\left(9 
H-\sqrt{6}\dot{\chi}\right)-\frac{1}{4}e^{-\sqrt{\frac{
2}{3}}\chi}+\frac{1}{2}=0.
\end{equation}
The corresponding effective dark-energy energy density and pressure 
(\ref{rhoDE}),(\ref{pDE}) become
\begin{equation}
  \label{rhoDE2}
 \rho_{DE}= 
 \frac{1}{2}\dot{\chi}^{2}-\frac{1}{8}e^{-2\sqrt{\frac{2}{3}}\chi}\f+\frac{1}{4}
e^{-\sqrt{\frac{2}{3}}\chi}\left(\f+\z\dot{\f}^{2} \right),
 \end{equation}
\begin{equation}
\label{pDE2}
p_{DE}=
 \frac{1}{2}\dot{\chi}^{2}+\frac{1}{8}e^{-2\sqrt{\frac{2}{3}}\chi}\f-\frac{1}{4}
e^{-\sqrt{\frac{2}{3}}\chi}\left(\f-\z\dot{\f}^{2} \right).
\end{equation}

In order to obtain the behaviour of the Hubble parameter, we first set 
$z=-1+a_0/a$, with the current value of the scale factor being set to $a_0=1$.
It is well know that the behaviour of the Hubble parameter in $\Lambda$CDM cosmology is given by
\begin{equation}
    H_{\Lambda CDM}(z)=H_0\sqrt{\Omega_{m_0}(1+z)^3+1-\Omega_{m_0}},
\end{equation}
where $H_0$ is the present value of the Hubble parameter and $\Omega_{m_0}$ is
the present value of matter density parameter defined as 
$\Omega_{m_0}=\frac{ \rho_m}{3H^2}$ in Planck units. We set $\Omega_{m_0}=0.31$ 
and 
$H_0=67.3 km/s/Mpc$. We then solve Eq. \eqref{Fr1rv}-\eqref{scalar2} numerically 
to obtain the solutions for the scale factor and hence for the Hubble 
parameter. In order to achieve this we set the initial conditions such that the 
evolution of $H(z)$ that we obtain  for 
$z=z_{CMB}\approx 1100$ coincides with $H_{\Lambda CDM}$, namely
$H(z\rightarrow z_{CMB})\approx H_{\Lambda CDM}$ while $H(z\rightarrow 
0)>H_{\Lambda CDM}(z\rightarrow 0)$.
For our present analysis we have one model parameter, i.e. $\zeta$, which 
determines the late-time deviation of the model from $\Lambda$CDM scenario.
\begin{figure}[H]
\begin{center}
\includegraphics[height=6.cm,width=8.5cm, clip=true]{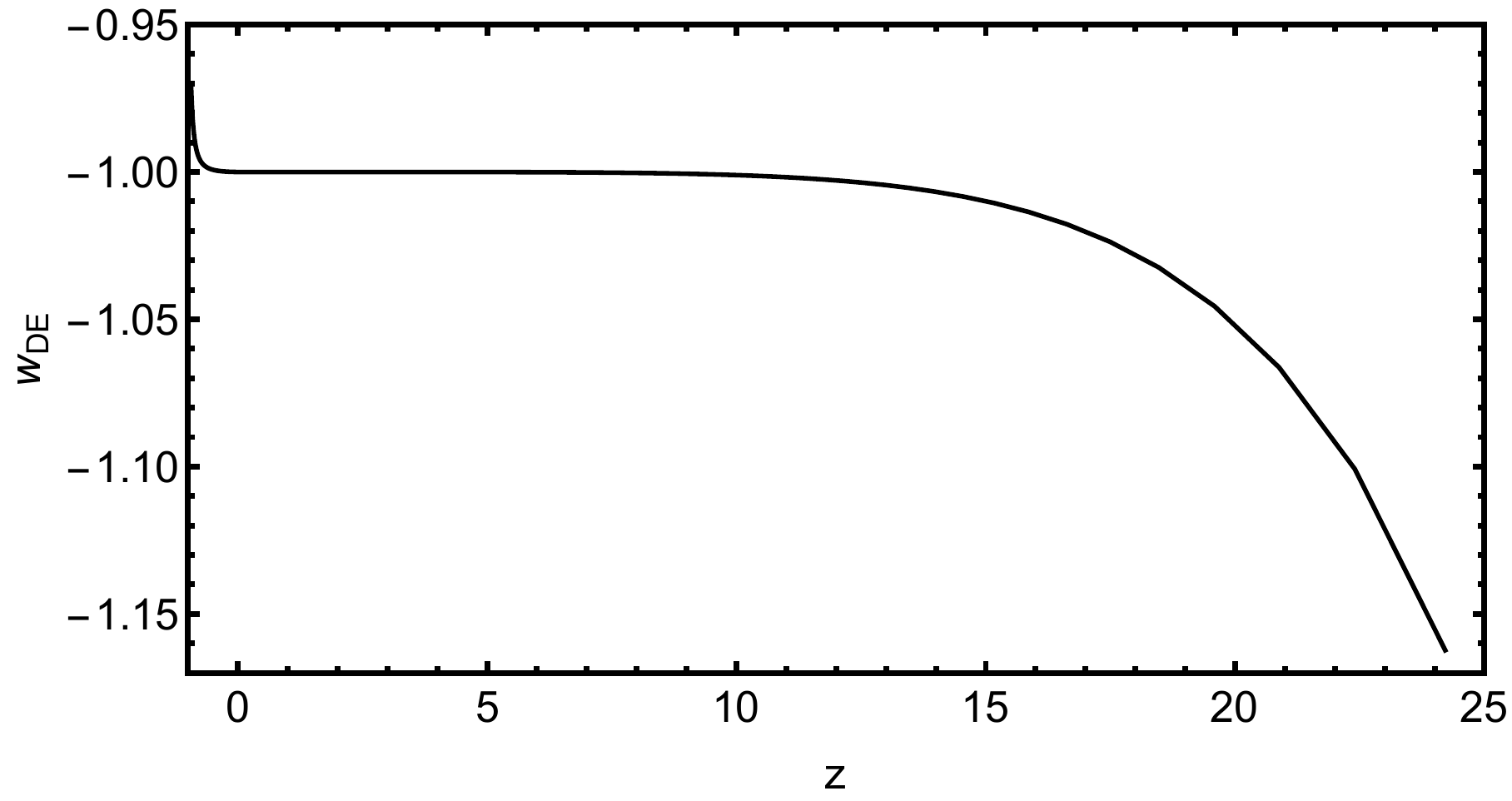}
\caption{{\it The effective dark-energy equation-of-state parameter $w_{DE}$ as 
a function of the redshift, for Model I for $\zeta=-10$  in  Planck  units.}}
\label{wde1}
\end{center}
\end{figure}

In Fig. \ref{wde1} we plot the evolution of the dark-energy equation-of-state 
parameter in terms of the redshift. As we can see from the figure,   
$w_{DE}<-1$ most of the time, thereby depicting phantom evolution which implies 
faster expansion. The phantom behavior is one of the mechanisms that  can
lead to the Hubble tension alleviation \cite{Yan:2019gbw,Heisenberg:2022lob} 
(see also the  discussion in \cite{Abdalla:2022yfr}), and as we will see in the 
following, this is exactly what happens.

\begin{figure}[H]
\begin{center}
\includegraphics[height=6.cm,width=8.5cm, clip=true]{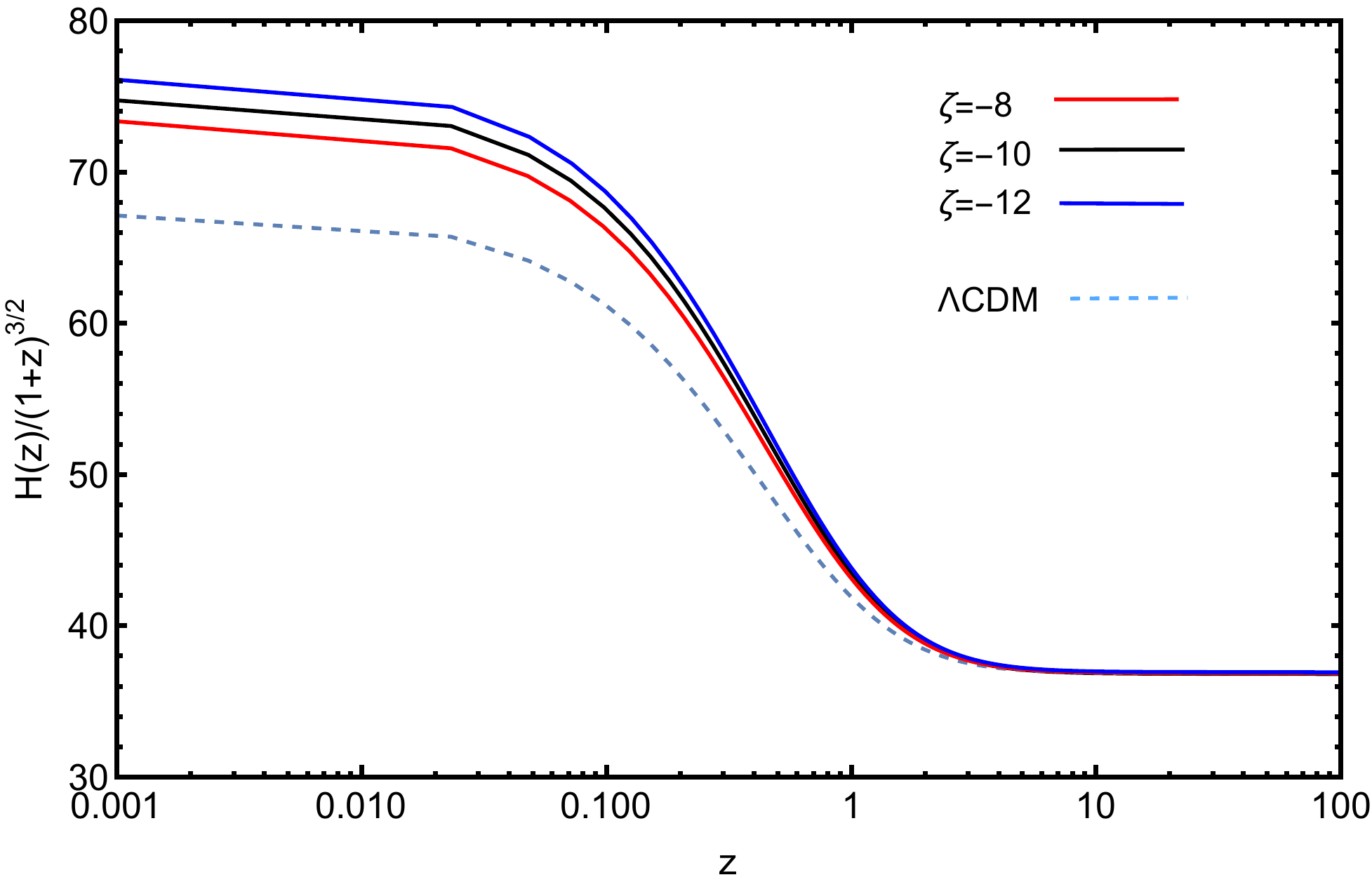}
\caption{{\it {The normalized combination $H(z)/(1+z)^{3/2}$  as a function of 
the redshift, for $\Lambda$CDM cosmology (blue dotted line), and for Model I 
for 
$\zeta=-12$ (solid blue 
line), for $\zeta=-10$ (solid black line), and for  $\zeta=-8$ (solid red 
line), in  Planck units.}}} 
\label{H1}
\end{center}
\end{figure}
In Fig. \ref{H1}, we present the normalised   combination 
$H(z)/(1+z)^{3/2}$ as a function of the redshift  for $\Lambda$CDM cosmology, 
and for Model I for different values of $\zeta$. Here we used     
$\zeta=-8,-10,-12$ in   Planck  units. We find that the present value of 
$H_0$ 
depends on the model parameter $\zeta$ as expected. For  $\zeta=-10$, the 
present value of the Hubble parameter is around $H_0\approx 74 km/s/Mpc$, which 
is consistent with the direct measurement   of the present Hubble parameter. 
 Values of $\zeta$ higher or lower than this lead to higher or lower values 
for 
$H_0$ respectively, and positive $\zeta$ corresponds to $H_0$ values lower 
than the value of $H_0$ in $\Lambda$CDM scenario, thus they are not relevant 
for our present analysis. Note that in natural units   $\zeta\sim-10$ 
corresponds to a typical value $\zeta^{1/4}\sim-10^{-19}$ GeV$^{-1}$.
Hence, such values are the ones 
that needed in order to bring $H_0$ from its $\Lambda$CDM value to the  
local-measurement value, in other words the magnitude 
and the sign of the modified gravity modification is phenomenologically 
determined by the distance of $H_0=67.3 
km/s/Mpc$ and  $H_0\approx 74 km/s/Mpc$.

\begin{figure}[H]
\begin{center}
\includegraphics[height=6.cm,width=8.5cm, clip=true]{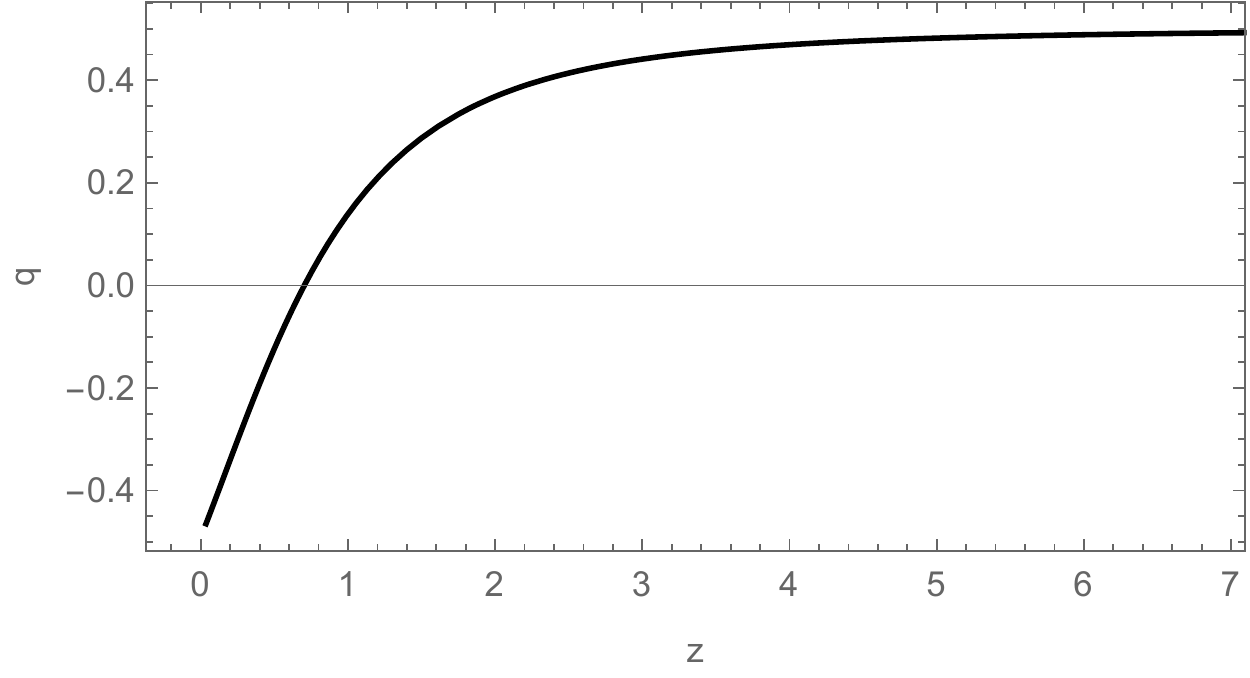}
\caption{{\it The deceleration parameter $q$ as a function of redshift $z$, for 
Model I with $\zeta=-10$ in  Planck  units.}}
\label{q1}
\end{center}
\end{figure}

For completeness, in Fig. \ref{q1}  we   depict the evolution of the 
deceleration  parameter $q\equiv -1-\dot{H}/H^2$ as a function of the  
redhsift. As we see, the redshift at which the transition from deceleration to 
acceleration occurs is around $z_{tr}=0.68$, in agreement with current 
observations.

 In summary, as we observe, there exist a range of the free model parameter 
$\zeta$ that is able to reproduce a Hubble function evolution that coincides 
with $\Lambda$CDM cosmology at high redshifts, but at late times it alleviates 
the $H_0$ tension. The reason that this happens is the fact that the 
effective dark-energy equation-of-state 
parameter exhibits a phantom behavior  (following the general requirements  of  
 \cite{Heisenberg:2022lob,Abdalla:2022yfr}).

\subsection{Model II}

As a next we consider the case where
\be
\mathcal{K}(\f,B)=\frac{1}{2}\f \quad \text{and} \quad \mathcal{G}(\f,B)=\xi B,
\ee
with $\xi$ the corresponding coupling constant  with
dimensions  $[M]^{-8}$.
Thus, the Friedmann equations  (\ref{FR1}),(\ref{FR2}) become
\bea
\label{Fr1mod2rv}
&&\!\!\!\!\!\!\!\!\!\!\!\!\!\!\!\!\!
3 
H^{2}-\rho_m-\frac{1}{2}\dot{\chi}^{2}+\frac{1}{8}e^{-2\sqrt{\frac{2}{3}}\chi}
\left(1-2 
e^{\sqrt{\frac{2}{3}}\chi} 
\right)\f\nonumber\\
&& \ \ \ \ \ \,
+\xi\dot{\f}^{3}\left(\sqrt{6}\dot{\chi}-6H 
\right)=0,
\eea
\bea
&&\!\!\!\!\!\!\!\!\!\!\!\!\!\!\!\!\!
3 H^{2}+2 
\dot{H}+p_m+\frac{1}{2}\dot{\chi}^{2}+\frac{1}{8}e^{-2\sqrt{\frac{2}{3}}\chi}
\left(1-2 
e^{\sqrt{\frac{2}{3}}\chi} 
\right)\f
\nonumber\\
&& \ \ \  \ \ \,
-\frac{1}{3}\xi\dot{\f}^{2}\left(\sqrt{6}\dot{\f}\dot{\chi}+6\ddot{\f} 
\right)=0,
\eea
while the two scalar field equations (\ref{chiequation}) and (\ref{phiequation}) 
read as
\bea
&&\!\!\!\!\!\!\!\!\!\!\!\!\!\!\!\!\!\!\!\!\!\!\!\!\!
\ddot{\chi}+3 H 
\dot{\chi}+\frac{1}{2\sqrt{6}}e^{-2\sqrt{\frac{2}{3}}\chi}\left(1-e^{\sqrt{\frac
{2}{
3}}\chi}\right)\f\nonumber\\
&&
-\sqrt{6}\,\xi\dot{\f}^{2}\left(H\dot{\f}+\ddot{\f} \right)=0,
\eea
\bea
&&\!\!\!\!\!\!\!\!\!\! 
\xi\dot{\f}\left\{2\left(-6H+\sqrt{6}\dot{\chi}\right)\ddot{\f}
\right.\nonumber\\
&&\left.
+\dot{\f}\left[-6\dot{H}
+3H\left(-6H+\sqrt{6}\dot{\chi}\right)+\sqrt{6}\ddot{\chi}\right]\right\}
\nonumber\\
&&\!\!\!\!\!\!\!\!\!
+\frac{1}{8}e^{
-2\sqrt{\frac{2}{ 3}}\chi}\left(1-2 e^{\sqrt{\frac{2}{3}}\chi} \right)=0.
\eea
Therefore, in this case the effective dark-energy energy density and pressure 
(\ref{rhoDE}),(\ref{pDE}) write as
\begin{eqnarray}
  \label{rhoDE3}
 &&\!\!\!\!\!\!\!\!\!\!\!\!\!\!\!\!\!\!
 \rho_{DE}= 
 \frac{1}{2}\dot{\chi}^{2}-\frac{1}{8}e^{-2\sqrt{\frac{2}{3}}\chi}\left(1-2 
e^{\sqrt{\frac{2}{3}}\chi} \right)\f
\nonumber\\
&&
-\xi\dot{\f}^{3}\left(\sqrt{6}\dot{\chi}-6H 
\right),
 \end{eqnarray}
\begin{eqnarray}
\label{pDE3}
 &&\!\!\!\!\!\!\!\!\!\!\!\!\!\!\!\!\!\!
p_{DE}=
\frac{1}{2}\dot{\chi}^{2}+\frac{1}{8}e^{-2\sqrt{\frac{2}{3}}\chi}\left(1-2 
e^{\sqrt{\frac{2}{3}}\chi} 
\right)\f
\nonumber\\
&&
-\frac{1}{3}\xi\dot{\f}^{2}\left(\sqrt{6}\dot{\f}\dot{\chi}+6\ddot{\f} 
\right).
\end{eqnarray}

\begin{figure}[H]
\begin{center}\hspace{-0.5cm}
\includegraphics[height=6.cm,width=8.2cm, clip=true]{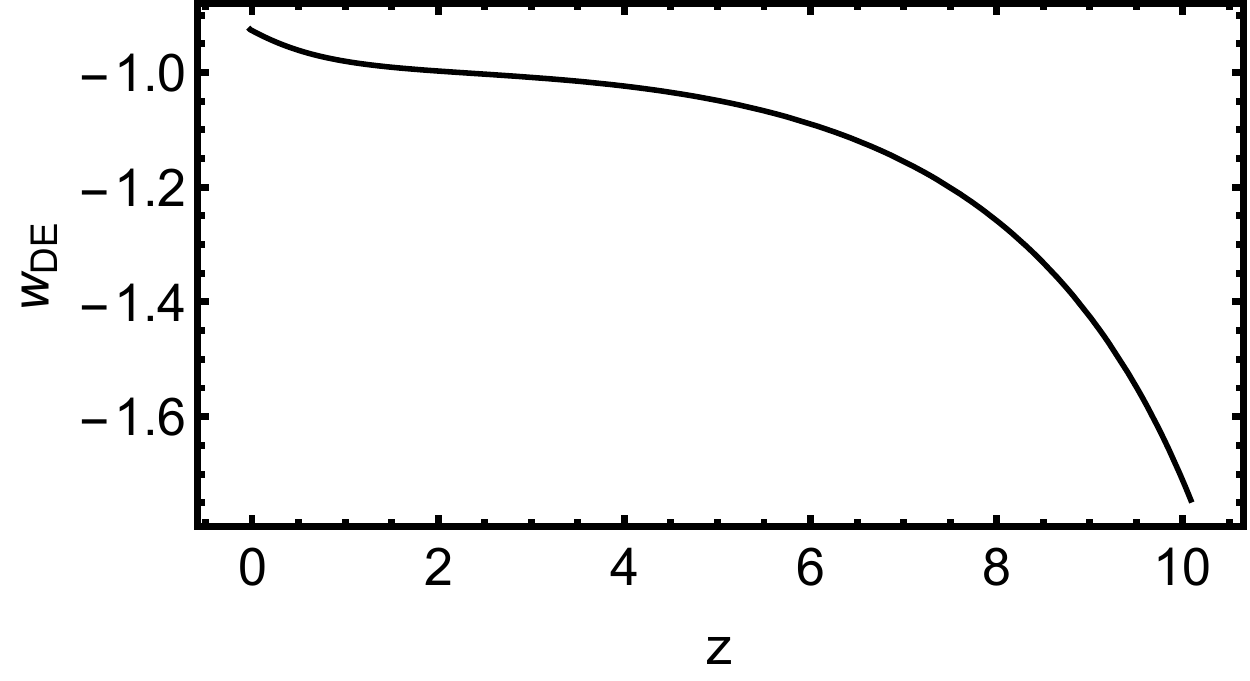}
\caption{{\it The effective dark-energy equation-of-state parameter $w_{DE}$ as 
a function of the redshift, for Model II for $\xi=-10$ in  Planck  units.}}
\label{wde2}
\end{center}
\end{figure}

Let us now proceed to the numerical investigation of the above equations.
Similarly to the previous Model I, we choose the initial conditions such 
that our scenario matches $\Lambda$CDM cosmology for $z\approx 1100$. 
In Fig. \ref{wde2}, we depict the evolution of the dark-energy 
equation-of-state parameter with the redshift. As in the case of the previous 
subsection, here   we also see that $w_{DE}<-1$ for most redshifts, thereby 
depicting phantom evolution, thus serving as a mechanism for Hubble tension 
alleviation. 

\begin{figure}[H]
\begin{center}
\includegraphics[height=6.cm,width=8.5cm, clip=true]{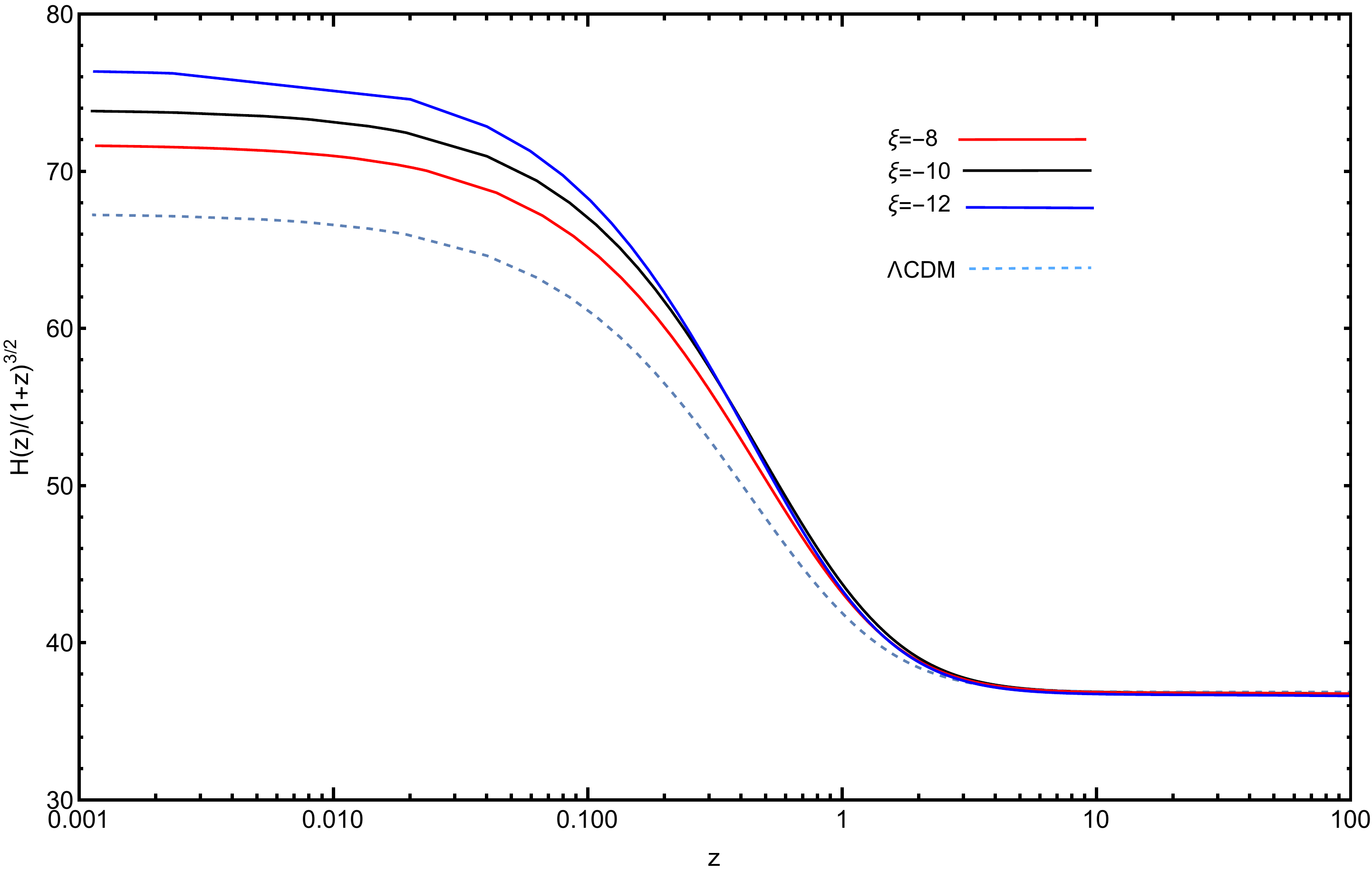}
\caption{{\it The normalized combination $H(z)/(1+z)^{3/2}$  as a function of 
the redshift, for $\Lambda$CDM cosmology (blue dotted line), and for Model II 
for 
$\xi=-12$ (solid blue 
line), for $\xi=-10$ (solid black line), and for $\xi=-8$ (solid red 
line), in  Planck  units. }}
\label{H2}
\end{center}
\end{figure}

In Fig. \ref{H2}, we present the normalised $H(z)/(1+z)^{3/2}$ as a function of 
the redshift  for $\Lambda$CDM cosmology, and for model II for different values 
of $\xi$, namely $\xi=-8,-10,-12$.  As expected, we find that the present 
Hubble value  $H_0$ depends on the model parameter $\xi$. Specifically,  for
  $\xi=-10$ it is around 
$H_0\approx 74 km/s/Mpc$, which is consistent with the directly measured value 
of the Hubble parameter. 
Values of $\xi$ higher or lower than this give higher or lower values for $H_0$ 
respectively. Note that in natural units   $\xi\sim-10$ 
corresponds to a typical value $\xi^{1/8}\sim-10^{-19}$ GeV$^{-1}$.
Hence, similarly to Model I above,  such values are the ones 
that are phenomenologically needed in order to bring $H_0$ from  $H_0=67.3 
km/s/Mpc$ to  $H_0\approx 74 km/s/Mpc$.

\begin{figure}[H]
\begin{center}
\includegraphics[height=6.cm,width=8.5cm, clip=true]{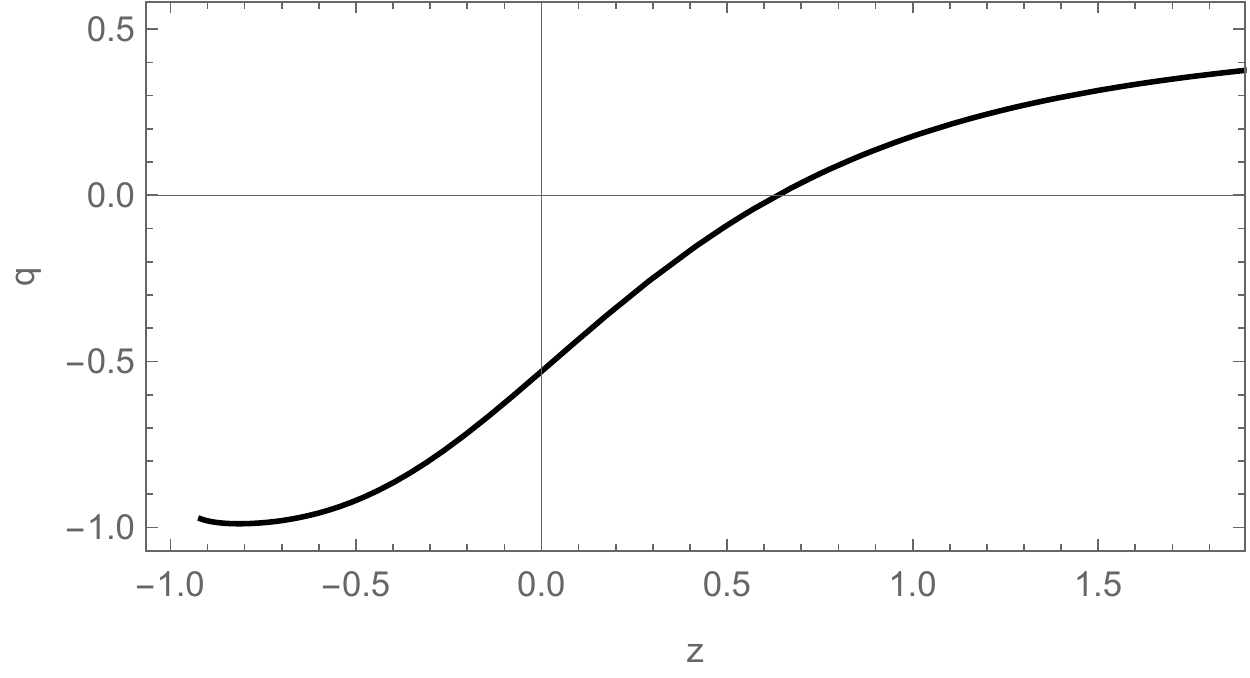}
\caption{{\it The deceleration parameter $q$ as a function of redshift $z$, for 
Model II with $\xi=-10$ in  Planck  units.}}
\label{q2}
\end{center}
\end{figure}

 In Fig. \ref{q2}, we depict the evolution of the deceleration parameter 
$q$ in terms of $z$. The transition redshift between 
deceleration and acceleration for this case is around $z_{tr}=0.65$, in 
agreement with current observations, too.

\begin{figure}[H]
\includegraphics[height=7cm,width=9.6cm, clip=true]{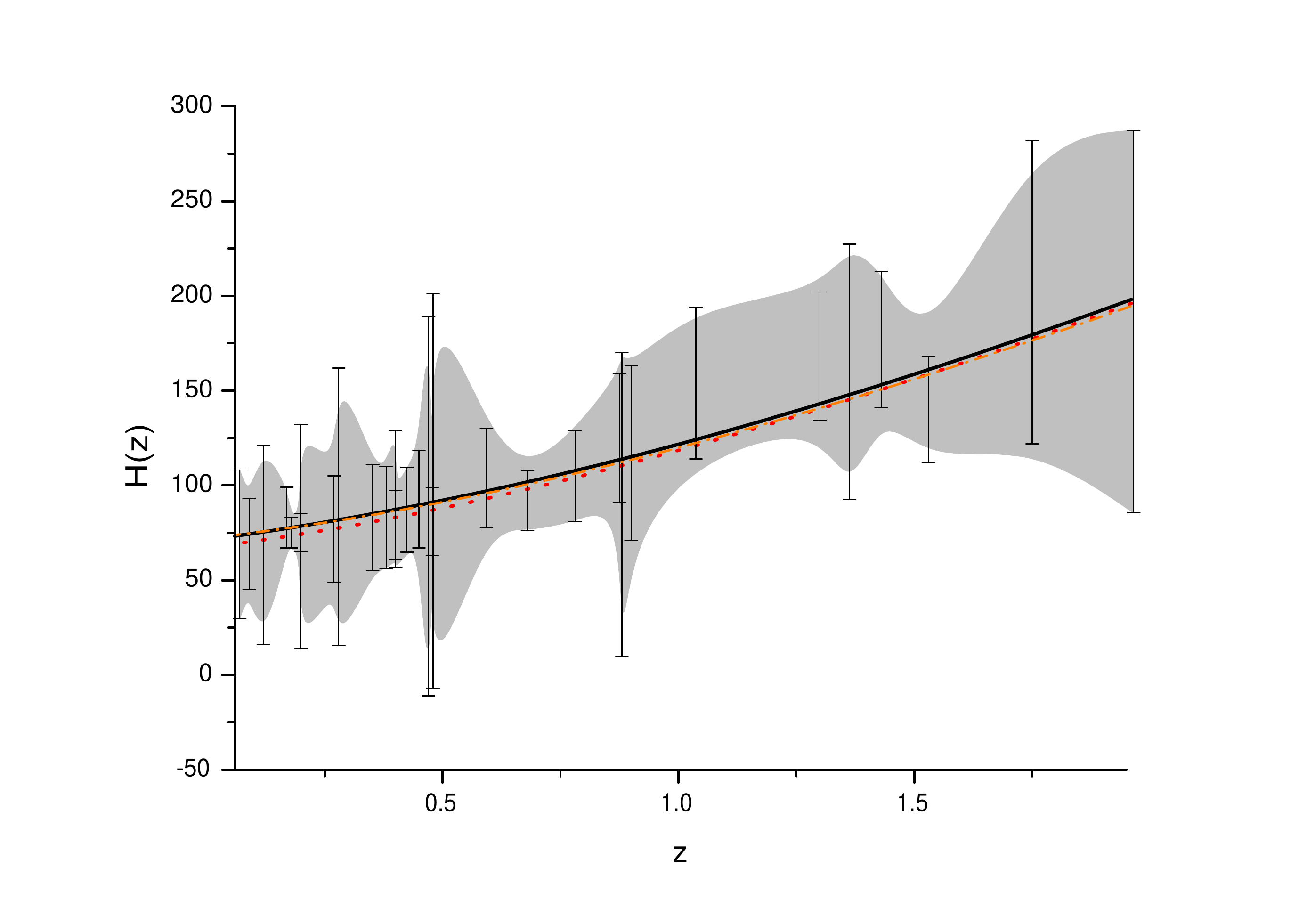}
\caption{{\it  The $H(z)$ in units of km/s/Mpc as a 
				function of the redshift, for $\Lambda$CDM scenario (red dotted 
line), 	for Model I with $\zeta=-10$   
(orange dashed-dotted), 		and 
				for Model II  with $\xi=-10$ (black solid line)  in  $H_0$  
units, on   top of the Cosmic Chronometers data 
points from \cite{Yu:2017iju} at $1\sigma$ confidence level.   We have imposed 
 $\Omega_{m_0}=0.31$.  	 }}
\label{CCdata}
\end{figure}

We close our analysis  by  confronting the two examined models with  Cosmic 
Chronometer (CC) cosmological data. This  
 datasets is based on the   $H(z)$ measurements  through the relative ages of 
 passively evolving galaxies and the corresponding estimation of 
 $dz/dt$   \cite{Jimenez:2001gg}. In Fig.   \ref{CCdata}   we   confront the 
  predicted $H(z)$ evolution of our models, alongside the one of 
$\Lambda$CDM 
scenario, with the $H(z)$ Cosmic Chronometer Data   \cite{Yu:2017iju} at 
$1\sigma$ confidence level. As we deduce, the agreement is very good, and the 
theoretical $H(z)$ evolution   lies within the  direct 
measurements of the $H(z)$ from the CC data.

\section{Conclusions}
\label{Conclusions}

New gravitational scalar-tensor theories  are novel modifications of gravity, 
consisting of a Lagrangian with the Ricci scalar and its first and second 
derivatives in a specific combination that makes the theory free of ghosts. Such 
constructions  propagate 2+2 degrees of freedom,  thus 
forming a subclass of bi-scalar extensions of general relativity.

In the present work we investigated the possibility of resolving the Hubble tension using these new gravitational scalar tensor theories.
 Considering a homogenoeus and isotropic background, we extracted the Friedmann 
equations, as well as the evolution equations of the new extra scalar degrees of 
freedom. We obtained an effective dark energy sector that consists of both 
extra scalar degrees of freedom. 

We then studied the cosmological behaviour of 
two specific models, imposing  as   initial conditions at high redshifts the 
coincidence of the behaviour of the Hubble function  with that predicted by 
$\Lambda$CDM cosmology. However, we showed that as time passes, the effect of 
bi-scalar modifications become important and thus at low redshifts the Hubble 
function acquires increased values in a controlled way. In particular, the 
present value of the Hubble parameter is sensitive to the choice of the model 
parameters.

In both models we showed that at high and intermediate redshifts 
the Hubble function behaves identically to that of $\Lambda$CDM scenario, 
however at low redshifts it acquires increased values, resulting to $H_0\approx 
74 km/s/Mpc$ for particular parameter choices. Hence, these new gravitational 
scalar tensor theories can alleviate the $H_0$ tension. The mechanism behind 
this behavior is the fact that the effective dark-energy equation-of-state 
parameter exhibits phantom behavior,  which implies 
faster expansion, and it is one of the sufficient theoretical requirements that 
are 
capable of alleviating the $H_0$ tension 
\cite{Heisenberg:2022lob,Abdalla:2022yfr} (although it is not a 
necessary requirement as we mention in the Introduction).
 Finally, we further confronted our models with Cosmic Chromometer data  and we 
found they are viable and in agreement with  observations.

It would be interesting to investigate what is the situation of the other 
famous tension, namely the $\sigma_8$ one (there seems to be a disagreement 
between the amount of matter clustering, quantified by $\sigma_8$, predicted by 
$\Lambda$CDM cosmology and the local measurements of the matter distribution 
\cite{Abdalla:2022yfr}) 
in the scenario at hand. In particular, a suggested solution for the $H_0$ 
tension does not guarantee an alleviation for the $\sigma_8$ one. There are 
models in which $H_0$ alleviation does impinge
positively on the $\sigma_8$ tension, such as the running vacuum ones
\cite{SolaPeracaula:2021gxi,Gomez-Valent:2018nib,Gomez-Valent:2017idt} or 
$f(T)$ gravity ones \cite{Yan:2019gbw,Saridakis:2023pzo}, however there are 
others in which  it leads to a worsening of the latter. That is why it is 
  necessary to perform a $\sigma_8$ analysis, too. Since 
such  an analysis requires the investigation of perturbations and the evolution 
of matter overdensity $\delta$, it is left for a separate project, however the 
obtained phantom behavior is expected to lead to an increase in the friction 
term in the Jeans equation for  $\delta$, which is qualitatively expected to 
lead to a smaller $\sigma_8$.

In conclusion, in this first work on the subject  we deduced that the  $H_0$ 
tension can be alleviated in the framework of new geometric gravitational 
theories. Definitely, the full verification of the above result requires a 
complete observational analysis, using  data from   Supernovae type Ia (SNIa), 
Baryonic Acoustic Oscillations (BAO), Redshift Space Distortion (RSD), and 
Cosmic Microwave Background (CMB)   observations.
Such a full and detailed observational confrontation, is left for a future 
project.

\begin{acknowledgments} 
M. P. is supported by the Basic Research  program 
of the National Technical University of Athens (NTUA, PEVE) 65232600-ACT-MTG: 
{\it{Alleviating Cosmological Tensions Through Modified Theories of Gravity}}.
The authors would like to acknowledge the contribution of the COST Action 
CA18108 ``Quantum Gravity Phenomenology in the multi-messenger approach'', as 
well as the contribution of the COST Action CA21136 ``Addressing observational 
tensions in  cosmology with systematics and fundamental physics (CosmoVerse)''.

\end{acknowledgments}


\begin{thebibliography}{99}
\bibitem{Aghanim:2018eyx} 
 N.~Aghanim {\it et al.} [Planck Collaboration],
  [\href{\arxiv/arXiv:1807.06209}{arXiv:1807.06209}].
 

\bibitem{Riess:2019cxk} 
  A.~G.~Riess, S.~Casertano, W.~Yuan, L.~M.~Macri and D.~Scolnic,
  Astrophys.\ J.\  {\bf 876}, 85 (2019)
  [\href{\arxiv/arXiv:1903.07603}{arXiv:1903.07603}].
  
 
 

\bibitem{Alam:2016hwk} 
  S.~Alam {\it et al.} [BOSS Collaboration],
  Mon.\ Not.\ Roy.\ Astron.\ Soc.\  {\bf 470}, 2617 (2017)
  [\href{\arxiv/arXiv:1607.03155}{arXiv:1607.03155 }].
  

\bibitem{Ata:2017dya} 
  M.~Ata {\it et al.},
  Mon.\ Not.\ Roy.\ Astron.\ Soc.\  {\bf 473}, 4773 (2018)
  [\href{\arxiv/arXiv:1705.06373}{arXiv:1705.06373}].
  
  

  
  \bibitem{Zarrouk:2018vwy} 
  P.~Zarrouk {\it et al.},
  Mon.\ Not.\ Roy.\ Astron.\ Soc.\  {\bf 477},  1639 (2018)
  [\href{\arxiv/arXiv:1801.03062}{arXiv:1801.03062}].
  
  
  
\bibitem{DiValentino:2020zio}
E.~Di Valentino, L.~A.~Anchordoqui, O.~Akarsu, Y.~Ali-Haimoud, L.~Amendola, 
N.~Arendse, M.~Asgari, M.~Ballardini, S.~Basilakos and E.~Battistelli, 
\textit{et al.}
Astropart. Phys. \textbf{131}, 102605 (2021)
  [\href{\arxiv/arXiv:2008.11284}{arXiv:2008.11284}].
  
\bibitem{DiValentino:2020vvd}
E.~Di Valentino, L.~A.~Anchordoqui, \"O.~Akarsu, Y.~Ali-Haimoud, L.~Amendola, 
N.~Arendse, M.~Asgari, M.~Ballardini, S.~Basilakos and E.~Battistelli, 
\textit{et al.}
Astropart. Phys. \textbf{131}, 102604 (2021)
  [\href{\arxiv/arXiv:2008.11285}{arXiv:2008.11285}].
 
\bibitem{Abdalla:2022yfr}
E.~Abdalla, G.~Franco Abell\'an, A.~Aboubrahim, A.~Agnello, O.~Akarsu, 
Y.~Akrami, G.~Alestas, D.~Aloni, L.~Amendola and L.~A.~Anchordoqui, \textit{et 
al.}
JHEAp \textbf{34}, 49-211 (2022)
  [\href{\arxiv/arXiv:2203.06142}{arXiv:2203.06142}].

 


\bibitem{DiValentino:2021izs}
E.~Di Valentino, O.~Mena, S.~Pan, L.~Visinelli, W.~Yang, A.~Melchiorri, 
D.~F.~Mota, A.~G.~Riess and J.~Silk,
Class. Quant. Grav. \textbf{38}, no.15, 153001 (2021)
[\href{\arxiv/arXiv:2103.01183}{arXiv:2103.01183}].
 

\bibitem{DiValentino:2015ola}
E.~Di Valentino, A.~Melchiorri and J.~Silk,
Phys. Rev. D \textbf{92}, no.12, 121302 (2015)
[\href{\arxiv/arXiv:1507.06646}{arXiv:1507.06646}].

\bibitem{Bernal:2016gxb}
J.~L.~Bernal, L.~Verde and A.~G.~Riess,
JCAP \textbf{10}, 019 (2016)
 \href{\arxiv/arXiv:1607.05617}{arXiv:1607.05617}].




\bibitem{Pan:2019gop}
S.~Pan, W.~Yang, E.~Di Valentino, E.~N.~Saridakis and S.~Chakraborty,
Phys. Rev. D \textbf{100}, 103520 (2019)
[\href{\arxiv/arXiv:1907.07540}{arXiv:1907.07540}].


\bibitem{Pan:2019jqh}
S.~Pan, W.~Yang, C.~Singha and E.~N.~Saridakis,
Phys. Rev. D \textbf{100}, 083539 (2019)
[\href{\arxiv/arXiv:1903.10969}{arXiv:1903.10969}].





\bibitem{Yang:2018prh}
W.~Yang, S.~Pan, E.~Di Valentino and E.~N.~Saridakis,
Universe \textbf{5}, 219 (2019)
[\href{\arxiv/arXiv:1811.06932}{arXiv:1811.06932}].



\bibitem{Yang:2018qmz}
W.~Yang, S.~Pan, E.~Di Valentino, E.~N.~Saridakis and S.~Chakraborty,
Phys. Rev. D \textbf{99}, 043543 (2019)
[\href{\arxiv/arXiv:1810.05141}{arXiv:1810.05141}].



 \bibitem{Kumar:2016zpg}
 S.~Kumar and R.~C.~Nunes,
 Phys. Rev. D \textbf{94}, no.12, 123511 (2016)
 [\href{\arxiv/arXiv:1608.02454}{arXiv:1608.02454}][astro-ph.CO].

\bibitem{DiValentino:2017iww}
E.~Di Valentino, A.~Melchiorri and O.~Mena,
Phys. Rev. D \textbf{96}, no.4, 043503 (2017)
[\href{\arxiv/arXiv:1704.08342}{arXiv:1704.08342}][astro-ph.CO].

\bibitem{DiValentino:2017oaw}
E.~Di Valentino, C.~Boehm, E.~Hivon and F.~R.~Bouchet,
Phys. Rev. D \textbf{97}, no.4, 043513 (2018)
[\href{\arxiv/arXiv:1710.02559}{arXiv:1710.02559}][astro-ph.CO].

 

\bibitem{Binder:2017lkj}
T.~Binder, M.~Gustafsson, A.~Kamada, S.~M.~R.~Sandner and M.~Wiesner,
Phys. Rev. D \textbf{97}, no.12, 123004 (2018)
[\href{\arxiv/arXiv:1712.01246}{arXiv:1712.01246}] [astro-ph.CO].

\bibitem{DiValentino:2017zyq}
E.~Di Valentino, A.~Melchiorri, E.~V.~Linder and J.~Silk,
Phys. Rev. D \textbf{96}, no.2, 023523 (2017)
[\href{\arxiv/arXiv:1704.00762}{arXiv:1704.00762}].

\bibitem{Yang:2018euj}
W.~Yang, S.~Pan, E.~Di Valentino, R.~C.~Nunes, S.~Vagnozzi and D.~F.~Mota,
JCAP \textbf{09}, 019 (2018)
[\href{\arxiv/arXiv:1805.08252}{arXiv:1805.08252}].

\bibitem{DEramo:2018vss}
F.~D'Eramo, R.~Z.~Ferreira, A.~Notari and J.~L.~Bernal,
JCAP \textbf{11}, 014 (2018)
[\href{\arxiv/arXiv:1808.07430}{arXiv:1808.07430}].

\bibitem{Poulin:2018cxd}
V.~Poulin, T.~L.~Smith, T.~Karwal and M.~Kamionkowski,
Phys. Rev. Lett. \textbf{122}, no.22, 221301 (2019)
[\href{\arxiv/arXiv:1811.04083}{arXiv:1811.04083}].

\bibitem{Shafieloo:2016bpk}
A.~Shafieloo, D.~K.~Hazra, V.~Sahni and A.~A.~Starobinsky,
Mon. Not. Roy. Astron. Soc. \textbf{473}, no.2, 2760-2770 (2018)
[\href{\arxiv/arXiv:1610.05192}{arXiv:1610.05192}] [astro-ph.CO].


 
 


\bibitem{Pandey:2019plg}
K.~L.~Pandey, T.~Karwal and S.~Das,
JCAP \textbf{07}, 026 (2020)
[\href{\arxiv/arXiv:1902.10636}{arXiv:1902.10636}].

\bibitem{Adhikari:2019fvb}
S.~Adhikari and D.~Huterer,
Phys. Dark Univ. \textbf{28}, 100539 (2020)
[\href{\arxiv/arXiv:1905.02278}{arXiv:1905.02278}].



\bibitem{Benisty:2019pxb}
D.~Benisty,
[\href{\arxiv/arXiv:1912.11124}{arXiv:1912.11124}].
 

\bibitem{Perez:2020cwa}
A.~Perez, D.~Sudarsky and E.~Wilson-Ewing,
Gen. Rel. Grav. \textbf{53}, no.1, 7 (2021)
[\href{\arxiv/arXiv:2001.07536}{arXiv:2001.07536}].

\bibitem{Pan:2020bur}
S.~Pan, W.~Yang and A.~Paliathanasis,
Mon. Not. Roy. Astron. Soc. \textbf{493}, no.3, 3114-3131 (2020)
[\href{\arxiv/arXiv:2002.03408}{arXiv:2002.03408}].

\bibitem{Benevento:2020fev}
G.~Benevento, W.~Hu and M.~Raveri,
Phys. Rev. D \textbf{101}, no.10, 103517 (2020)
[\href{\arxiv/arXiv:2002.11707}{arXiv:2002.11707}].



\bibitem{Banerjee:2020xcn}
A.~Banerjee, H.~Cai, L.~Heisenberg, E.~\'O.~Colg\'ain, M.~M.~Sheikh-Jabbari and 
T.~Yang,
Phys. Rev. D \textbf{103}, no.8, L081305 (2021)
[\href{\arxiv/arXiv:2006.00244}{arXiv:2006.00244}].
 



\bibitem{Elizalde:2020mfs}
E.~Elizalde, M.~Khurshudyan, S.~D.~Odintsov and R.~Myrzakulov,
Phys. Rev. D \textbf{102}, no.12, 123501 (2020)
[\href{\arxiv/arXiv:2006.01879}{arXiv:2006.01879}].


 


\bibitem{Alvarez:2020xmk}
P.~D.~Alvarez, B.~Koch, C.~Laporte and \'A.~Rinc\'on,
JCAP \textbf{06}, 019 (2021)
[\href{\arxiv/arXiv:2009.02311}{arXiv:2009.02311}].

\bibitem{DeFelice:2020cpt}
A.~De Felice, S.~Mukohyama and M.~C.~Pookkillath,
Phys. Lett. B \textbf{816}, 136201 (2021)
[\href{\arxiv/arXiv:2009.08718}{arXiv:2009.08718}].

\bibitem{Haridasu:2020pms}
B.~S.~Haridasu, M.~Viel and N.~Vittorio,
Phys. Rev. D \textbf{103}, no.6, 063539 (2021)
[\href{\arxiv/arXiv:2012.10324}{arXiv:2012.10324}].


\bibitem{Seto:2021xua}
O.~Seto and Y.~Toda,
Phys. Rev. D \textbf{103}, no.12, 123501 (2021)
[\href{\arxiv/arXiv:2101.03740}{arXiv:2101.03740}].


\bibitem{Bernal:2021yli}
J.~L.~Bernal, L.~Verde, R.~Jimenez, M.~Kamionkowski, D.~Valcin and 
B.~D.~Wandelt,
Phys. Rev. D \textbf{103}, no.10, 103533 (2021)
[\href{\arxiv/arXiv:2102.05066}{arXiv:2102.05066}].


\bibitem{Alestas:2021xes}
G.~Alestas and L.~Perivolaropoulos,
Mon. Not. Roy. Astron. Soc. \textbf{504}, no.3, 3956-3962 (2021)
[\href{\arxiv/arXiv:2103.04045}{arXiv:2103.04045}].

\bibitem{Elizalde:2021kmo}
E.~Elizalde, J.~Gluza and M.~Khurshudyan,
[\href{\arxiv/arXiv:2104.01077}{arXiv:2104.01077}].

\bibitem{Krishnan:2021dyb}
C.~Krishnan, R.~Mohayaee, E.~\'O.~Colg\'ain, M.~M.~Sheikh-Jabbari and L.~Yin,
Class. Quant. Grav. \textbf{38}, no.18, 184001 (2021)
 [\href{\arxiv/arXiv:2105.09790}{arXiv:2105.09790}].

 



\bibitem{Theodoropoulos:2021hkk}
A.~Theodoropoulos and L.~Perivolaropoulos,
the 
Universe \textbf{7}, no.8, 300 (2021)
[\href{\arxiv/arXiv:2109.06256}{arXiv:2109.06256}].



\bibitem{Anagnostopoulos:2019miu}
F.~K.~Anagnostopoulos, S.~Basilakos and E.~N.~Saridakis,
Phys. Rev. D \textbf{100}, 083517 (2019)
[\href{\arxiv/arXiv:1907.07533}{arXiv:1907.07533}].

\bibitem{El-Zant:2018bsc}
A.~El-Zant, W.~El Hanafy and S.~Elgammal,
Astrophys. J. \textbf{871}, 210 (2019)
[\href{\arxiv/arXiv:1809.09390}{arXiv:1809.09390}].



\bibitem{Braglia:2020auw}
M.~Braglia, M.~Ballardini, F.~Finelli and K.~Koyama,
[\href{\arxiv/arXiv:2011.12934}{arXiv:2011.12934 }].



\bibitem{Abadi:2020hbr}
T.~Abadi and E.~D.~Kovetz,
[\href{\arxiv/arXiv:2011.13853}{arXiv:2011.13853} ].




\bibitem{Cai:2019bdh}
Y.~F.~Cai, M.~Khurshudyan and E.~N.~Saridakis,
Astrophys. J. \textbf{888}, 62 (2020)
[\href{\arxiv/arXiv:1907.10813}{arXiv:1907.10813}].



\bibitem{Escamilla-Rivera:2019ulu}
C.~Escamilla-Rivera and J.~Levi Said,
Class. Quant. Grav. \textbf{37}, 165002 (2020)
[\href{\arxiv/arXiv:1909.10328}{arXiv:1909.10328}].


\bibitem{Barker:2020gcp}
W.~E.~V.~Barker, A.~N.~Lasenby, M.~P.~Hobson and W.~J.~Handley,
Phys. Rev. D \textbf{102}, 024048 (2020)
[\href{\arxiv/arXiv:2003.02690}{arXiv:2003.02690}].



\bibitem{Wang:2020zfv}
D.~Wang and D.~Mota,
Phys. Rev. D \textbf{102}, 063530 (2020)
[\href{\arxiv/arXiv:2003.10095}{arXiv:2003.10095}].


\bibitem{Ballardini:2020iws}
M.~Ballardini, M.~Braglia, F.~Finelli, D.~Paoletti, A.~A.~Starobinsky and 
C.~Umilt\`a,
JCAP \textbf{10}, 044 (2020)
[\href{\arxiv/arXiv:2004.14349}{arXiv:2004.14349}].


\bibitem{LinaresCedeno:2020uxx}
F.~X.~Linares Cede\~no and U.~Nucamendi,
[\href{\arxiv/arXiv:2009.10268}{arXiv:2009.10268}].



\bibitem{Basilakos:2018arq}
S.~Basilakos, S.~Nesseris, F.~K.~Anagnostopoulos and E.~N.~Saridakis,
JCAP \textbf{08}, 008 (2018)
[\href{\arxiv/arXiv:1803.09278}{arXiv:1803.09278}].


\bibitem{Adil:2021zxp}
S.~A.~Adil, M.~R.~Gangopadhyay, M.~Sami and M.~K.~Sharma,
[\href{\arxiv/arXiv:2106.03093}{arXiv:2106.03093}].

\bibitem{Odintsov:2020qzd}
S.~D.~Odintsov, D.~S.~C.~G\'omez and G.~S.~Sharov,
Nucl. Phys. B \textbf{966}, 115377 (2021)
[\href{\arxiv/arXiv:2011.03957}{arXiv:2011.03957}].

\bibitem{Nojiri:2022ski}
S.~Nojiri, S.~D.~Odintsov and V.~K.~Oikonomou,
Nucl. Phys. B \textbf{980} (2022), 115850
[\href{\arxiv/arXiv:2205.11681}{arXiv:2205.11681}].

 

\bibitem{Nojiri:2019fft}
S.~Nojiri, S.~D.~Odintsov and V.~K.~Oikonomou,
Phys. Dark Univ. \textbf{29} (2020), 100602
[\href{\arxiv/arXiv:1912.13128}{arXiv:1912.13128}].

 
 
 

\bibitem{CANTATA:2021ktz}
E.~N.~Saridakis \textit{et al.} [CANTATA],
Springer, 2021,
[\href{\arxiv/arXiv:1912.13128}{arXiv:1912.13128}].

 
 

 
  


\bibitem{DiValentino:2019jae}
E.~Di Valentino, A.~Melchiorri, O.~Mena and S.~Vagnozzi,
Phys. Rev. D \textbf{101}, 063502 (2020)
[\href{\arxiv/arXiv:1910.09853}{arXiv:1910.09853}].

 
 
\bibitem{Vagnozzi:2019ezj}
S.~Vagnozzi,
Phys. Rev. D \textbf{102}, 023518 (2020)
[\href{\arxiv/arXiv:1907.07569}{arXiv:1907.07569}].


\bibitem{Hu:2015rva}
B.~Hu and M.~Raveri,
Phys. Rev. D \textbf{91}, no.12, 123515 (2015)
[\href{\arxiv/arXiv:1502.06599}{arXiv:1502.06599}].

\bibitem{Khosravi:2017hfi}
N.~Khosravi, S.~Baghram, N.~Afshordi and N.~Altamirano,
Phys. Rev. D \textbf{99}, no.10, 103526 (2019)
[\href{\arxiv/arXiv:1710.09366}{arXiv:1710.09366}].

\bibitem{Belgacem:2017cqo}
E.~Belgacem, Y.~Dirian, S.~Foffa and M.~Maggiore,
JCAP \textbf{03}, 002 (2018)
[\href{\arxiv/arXiv:1712.07066}{arXiv:1712.07066}] [hep-th].

\bibitem{Nunes:2018xbm}
R.~C.~Nunes,
JCAP \textbf{05}, 052 (2018)
[\href{\arxiv/arXiv:1802.02281}{arXiv:1802.02281}].

\bibitem{Lin:2018nxe}
M.~X.~Lin, M.~Raveri and W.~Hu,
Phys. Rev. D \textbf{99}, no.4, 043514 (2019)
[\href{\arxiv/arXiv:1810.02333}{arXiv:1810.02333}].






\bibitem{DAgostino:2020dhv}
R.~D'Agostino and R.~C.~Nunes,
Phys. Rev. D \textbf{101}, no.10, 103505 (2020)
[\href{\arxiv/arXiv:2002.06381}{arXiv:2002.06381}].




\bibitem{Anagnostopoulos:2020lec}
F.~K.~Anagnostopoulos, S.~Basilakos and E.~N.~Saridakis,
[\href{\arxiv/arXiv:2012.06524}{arXiv:2012.06524}].



\bibitem{Capozziello:2020nyq}
S.~Capozziello, M.~Benetti and A.~D.~A.~M.~Spallicci,
Found. Phys. \textbf{50}, no.9, 893-899 (2020)
[\href{\arxiv/arXiv:2007.00462}{arXiv:2007.00462}].

\bibitem{Saridakis:2019qwt}
E.~N.~Saridakis, S.~Myrzakul, K.~Myrzakulov and K.~Yerzhanov,
Phys. Rev. D \textbf{102}, no.2, 023525 (2020)
[\href{\arxiv/arXiv:1912.03882}{arXiv:1912.03882}].


\bibitem{daSilva:2020bdc}
W.~J.~C.~da Silva and R.~Silva,
Eur. Phys. J. Plus \textbf{136}, no.5, 543 (2021)
[\href{\arxiv/arXiv:2011.09520}{arXiv:2011.09520}].




\bibitem{Sola:2017znb}
J.~Sol\`a, A.~G\'omez-Valent and J.~de Cruz P\'erez,
Phys. Lett. B \textbf{774}, 317-324 (2017)
[\href{\arxiv/arXiv:1705.06723}{arXiv:1705.06723}] [astro-ph.CO].


\bibitem{SolaPeracaula:2021gxi}
J.~Sol\`a Peracaula, A.~G\'omez-Valent, J.~de Cruz Perez and C.~Moreno-Pulido,
EPL \textbf{134}, no.1, 19001 (2021)
[\href{\arxiv/arXiv:2102.12758}{arXiv:2102.12758}].

  
\bibitem{Moreno-Pulido:2022phq}
C.~Moreno-Pulido and J.~Sola Peracaula,
Eur. Phys. J. C \textbf{82}, no.6, 551 (2022)
[\href{\arxiv/arXiv:2201.05827}{arXiv:2201.05827}].

 
 

\bibitem{Moreno-Pulido:2020anb}
C.~Moreno-Pulido and J.~Sola,
Eur. Phys. J. C \textbf{80}, no.8, 692 (2020)
[\href{\arxiv/arXiv:2005.03164}{arXiv:2005.03164}].

 
   

\bibitem{SolaPeracaula:2022hpd}
J.~Sola Peracaula,
Phil. Trans. Roy. Soc. Lond. A \textbf{380}, 20210182 (2022)
[\href{\arxiv/arXiv:2203.13757}{arXiv:2203.13757}].


\bibitem{Naruko:2015zze}
  A.~Naruko, D.~Yoshida and S.~Mukohyama,
  Class.\ Quant.\ Grav.\  {\bf 33}, no. 9, 09LT01 (2016)
[\href{\arxiv/arXiv:1512.06977}{arXiv:1512.06977}].
 
   
   

\bibitem{Saridakis:2016ahq} 
  E.~N.~Saridakis and M.~Tsoukalas,
  Phys.\ Rev.\ D {\bf 93}, no. 12, 124032 (2016)
  [\href{\arxiv/arXiv:1601.06734}{arXiv:1601.06734}].
  

\bibitem{Saridakis:2018fth}
E.~N.~Saridakis, S.~Banerjee and R.~Myrzakulov,
Phys. Rev. D \textbf{98}, no.6, 063513 (2018)
  [\href{\arxiv/arXiv:1807.00346}{arXiv:1807.00346}].
 
  
\bibitem{Horndeski:1974wa}
G.~W.~Horndeski,
\href{\doi/doi:10.1007/BF01807638}{Int. J. Theor. Phys. \textbf{10}, 363-384} 
(1974).

\bibitem{Nicolis:2008in}
  A.~Nicolis, R.~Rattazzi and E.~Trincherini,
  Phys.\ Rev.\ D {\bf 79} (2009) 064036
  [\href{\arxiv/arXiv:0811.2197} {arXiv:0811.2197}].
  
 
  


\bibitem{Deffayet:2011gz}
C.~Deffayet, X.~Gao, D.~A.~Steer and G.~Zahariade,
Phys. Rev. D \textbf{84}, 064039 (2011)
[\href{\arxiv/arXiv:1103.3260} {arXiv:1103.3260}].


\bibitem{Gleyzes:2014dya} 
  J.~Gleyzes, D.~Langlois, F.~Piazza and F.~Vernizzi,
  Phys.\ Rev.\ Lett.\  {\bf 114}, no. 21, 211101 (2015)
  [\href{\arxiv/arXiv:1404.6495} {arXiv:1404.6495}].

  
\bibitem{Yan:2019gbw}
S.~F.~Yan, P.~Zhang, J.~W.~Chen, X.~Z.~Zhang, Y.~F.~Cai and E.~N.~Saridakis,
Phys. Rev. D \textbf{101}, no.12, 121301 (2020)
[\href{\arxiv/arXiv:1909.06388}{arXiv:1909.06388}].

\bibitem{Heisenberg:2022lob}
L.~Heisenberg, H.~Villarrubia-Rojo and J.~Zosso,
[\href{\arxiv/arXiv:2201.11623}{arXiv:2201.11623}].

 


\bibitem{Jimenez:2001gg}
R.~Jimenez and A.~Loeb,
Astrophys. J. \textbf{573}, 37-42 (2002)
[\href{\arxiv/arXiv:astro-ph/0106145}{arXiv:astro-ph/0106145}].

  


\bibitem{Yu:2017iju}
H.~Yu, B.~Ratra and F.~Y.~Wang,
Astrophys. J. \textbf{856}, no.1, 3 (2018)
[\href{\arxiv/arXiv:1711.03437}{arXiv:1711.03437}].




\bibitem{Gomez-Valent:2018nib}
A.~G\'omez-Valent and J.~Sol\`a Peracaula,
Mon. Not. Roy. Astron. Soc. \textbf{478}, no.1, 126-145 (2018)
[\href{\arxiv/arXiv:1801.08501}{arXiv:1801.08501}].

 

\bibitem{Gomez-Valent:2017idt}
A.~Gomez-Valent and J.~Sola,
EPL \textbf{120}, no.3, 39001 (2017)
[\href{\arxiv/arXiv:1711.00692}{arXiv:1711.00692}].

 
\bibitem{Saridakis:2023pzo}
E.~N.~Saridakis,
[\href{\arxiv/arXiv:2301.06881}{arXiv:2301.06881}].

 


\end{thebibliography}
\end{document}